# Influence of Cu deposition potential on the giant magnetoresistance and surface roughness of electrodeposited Ni-Co/Cu multilayers

B.G. Tóth[a,*], L. Péter[a,1], J. Dégi[a], Á. Révész[b], D. Oszetzky[a], G. Molnár[c] and I. Bakonyi[a]

[a]*Institute for Solid State Physics and Optics, Wigner Research Centre for Phyisics, Hungarian Academy of Sciences.*
*H-1525 Budapest, P.O.B. 49, Hungary*

[b]*Department of Materials Physics, Eötvös University.*
*H-1518 Budapest, P.O.B. 32, Hungary*

[c]*Institute for Technical Physics and Materials Science, Research Centre for Natural Sciences, Hungarian Academy of Sciences*
*H-1525 Budapest, P.O.B. 49, Hungary*

**Abstract** ─ It has been shown previously for electrodeposited Co/Cu multilayers that the single-bath electrodeposition process can be optimized from an electrochemical point of view in order to avoid unwanted Co dissolution and incorporation of Co in the non-magnetic layer during the Cu deposition pulse. In the present work, electrodeposition of Ni-Co/Cu multilayers has been studied to clarify if the same optimization method is appropriate when two magnetic elements are present and if this potential results in the largest giant magnetoresistance (GMR) for the particular alloy system studied. For this purpose, several Ni-Co/Cu multilayers were prepared by varying the deposition potential of the Cu layer. The composition analysis of the deposits showed that the Ni:Co ratio exhibits a minimum as a function of the Cu deposition potential, which can be explained by considering both the dissolution of Co and the mass transport of the reactants. Both the saturation GMR value and the intensity of the satellite peaks in the X-ray diffractograms were highly correlated with the resulting surface roughness of the deposits which was strongly varying with the Cu deposition potential. Higher GMR values, lower saturation fields and more perfect multilayer structure were observed for sufficiently positive Cu deposition potentials only which enabled a partial Co dissolution resulting in a reduced surface roughness. The results draw attention to the complexity of the optimization procedure of the deposition of multilayers with several alloying components.

---

[*]Corresponding author. E-mail: toth.bence@wigner.mta.hu
[1] ISE active member

# 1. Introduction

The giant magnetoresistance (GMR) effect has been widely studied in various electrodeposited (ED) nanoscale ferromagnetic/non-magnetic (FM/NM) metallic multilayers over the last two decades (see the comprehensive review of this field in Ref. 1). A lot of efforts have been devoted to clarify the underlying electrochemical processes influencing the layer formation in ED multilayers. It was established that these processes depend on the deposition control mode [1,2].

When both the magnetic and non-magnetic layers are deposited under galvanostatic (G) control (G/G mode), a significant exchange reaction takes usually place [3-7]. During the deposition pulse of the more noble element (here, Cu) with a cathodic current, the electrode potential is ill-defined. The Cu deposition current density must be below the diffusion limited current density in order to avoid the contamination of the Cu layers with Co or Ni atoms. Therefore, the $Cu^{2+}$ ions not used in the galvanic process can oxidize the previously deposited magnetic metal. For the processes of the exchange reaction, a charge balance is maintained, i.e., $j_{exch}(Co) + j_{exch}(Ni) + j_{exch}(Cu) = 0$ at all times whereby generally the relation $|j_{exch}(Co)| \gg |j_{exch}(Ni)|$ holds. As a consequence, the Cu layer thickness will be larger than the preset nominal value. It was shown for ED Co/Cu multilayers [8,9] that the excess Cu layer thickness due to the exchange reaction can be as high as 1.4 nm. Evidently, a corresponding reduction of the magnetic layer thickness occurs.

The exchange reaction can be completely suppressed only if the more noble component (Cu) is deposited under potentiostatic (P) control [1,2] whereby the deposition mode of the less noble components (Co and/or Ni) can be either G or P. However, even in the P/P or G/P modes, the Cu deposition potential $E_{Cu}$ should be optimized in order to avoid any unwanted electrochemical reactions during the Cu deposition cycle [1,2,10,11]. Namely, if $E_{Cu}$ is more positive than the electrochemically optimum value $E_{Cu}^{EC}$, then a dissolution of the less noble magnetic atoms (Co and Ni) of the previously deposited magnetic layer occurs during the Cu deposition pulse. This dissolution takes place as long as the magnetic atoms are subjected to the electrolyte, i.e., not covered by the depositing Cu atoms. The relative dissolution preference of Co and Ni and the accompanying layer thickness changes are the same as for the exchange reaction in the G/G mode. On the other hand, if $E_{\text{Cu}}$ is more negative than $E_{Cu}^{EC}$, then the deposition of the less noble magnetic atoms proceeds along with Cu and, thus, the non-magnetic layer will be contaminated with magnetic atoms.

The establishment of an expected optimum for the Cu deposition potential is often based on an inspection of the cyclic voltammogram (CV) curves of the electrolyte used. It has been pointed out [10], however, that this method can be misleading since the CV curves depend substantially on several cycling parameters, especially in the potential range of interest for Cu layer deposition. It was suggested [8,10], instead, that an analysis of the current transients during the Cu deposition cycle under pulse plating conditions identical to the preparation of the multilayer is a proper way of establishing $E_{Cu}^{EC}$. This optimization procedure was successfully applied also for preparing FeCoNi/Cu multilayers with high GMR sensitivity [12]. Other methods have also been proposed [13,14] that yield a fairly reliable optimum Cu deposition potential from the electrochemical point of view. The suppression of the dissolution of the magnetic metal(s) with a corrosion inhibitor or surface blocking agent is not a viable opion when the target parameter is the magnetoresistance. The application of additives, as it was shown in the early years of ED multilayer research [13], has a detrimental effect on the GMR properties due to the grain refinement and the increase of the zero-field resistivity.

However, no systematic studies have been carried out to establish to what extent the electrochemically optimized Cu deposition potential corresponds to the largest possible magnetoresistance for a given element combination in the multilayers. It has been reported for ED Co-Cu/Cu multilayers [8] that fairly large GMR could be achieved even at Cu deposition potentials significantly different from $E_{Cu}^{EC}$, specifically at more positive potentials. Although several studies of the GMR were performed on ED multilayers with various Cu deposition potentials [1],

Chassaing et al. [15] performed only a systematic study on the effect of the Cu deposition potential between -0.25 V and -0.60 V vs. SCE. It was found in the latter study that the GMR of ED Co/Cu multilayers was the largest at about -0.35 V, although a significant GMR was measured at 5 K only.

It was, therefore, the main aim of the present work to systematically study, after the establishment of $E_{Cu}^{EC}$, the influence of Cu deposition potential around $E_{Cu}^{EC}$ on the GMR in order to see to what extent this electrochemically optimized potential corresponds to a possible maximum of GMR for the given system. For this purpose, we have chosen the Ni-Co/Cu multilayer system with approximately equal concentrations of Co and Ni in the magnetic layer. For a better understanding of the observed variation of the magnetoresistance with $E_{Cu}$, measurements of the overall chemical composition, X-ray diffraction (XRD) patterns and surface roughness have also been performed on these multilayers.

## 2. Experimental

*2.1 Sample preparation and characterization*

The Ni-Co/Cu multilayers were deposited from an aqueous electrolyte which was based, by the addition of $CuSO_4$, on a bath formerly elaborated for the preparation of Ni-Co alloys [16]. The composition used for multilayer deposition was 0.010 mol/ℓ $CuSO_4$, 0.10 mol/ℓ $Na_2SO_4$, 0.25 mol/ℓ $H_3BO_3$, 0.25 mol/ℓ $H_2NSO_3H$, 0.703 mol/ℓ $NiSO_4$ and 0.037 mol/ℓ $CoSO_4$. The pH was set to 3.25 by adding NaOH to the solution. The choice of this pH value was based on some preliminary experiments to get appropriate deposition conditions. The $Ni^{2+}$ and $Co^{2+}$ ionic concentrations in the electrolyte were chosen to get approximately equal amounts of Ni and Co in the magnetic layer of the deposited multilayer.

The Ni-Co/Cu multilayers were deposited on a [100]-oriented, 0.26 mm thick silicon wafer covered with a 5 nm Cr and a 20 nm Cu layer by evaporation. The purpose of the Cr layer was to assure adhesion and the Cu-layer was used to provide the electrical conductivity of the cathode surface.

The deposition was performed in a tubular cell of 8 mm x 20 mm cross section with an upward looking cathode at the bottom of the cell [8,17]. The multilayer preparation was carried out by using a G/P pulse combination [1,2,8]. For the deposition of the magnetic layer, G mode was used at -35.0 mA/cm$^2$ current density. At this current density, the amount of Cu incorporated in the magnetic layer from the dilute Cu$^{2+}$ solution does not deteriorate the magnetic and transport properties of the Ni-Co layer. For the Cu-layer, P mode was applied and a saturated calomel electrode (SCE) served as reference. Based on the optimization method described in Ref. 10, it was found that $E_{Cu}^{EC}$ = -0.585 V is the electrochemically optimized Cu deposition potential for this particular bath with about -0.53 mA/cm$^2$ diffusion-limited current density. By controlling the deposition time in the G mode (magnetic layer deposition) and the charge driven through the cell in the P mode (non-magnetic layer deposition), the individual layer thicknesses could be set to a predetermined value by using Faraday's law. For Cu deposition, 100 % current efficiency was assumed which is the commonly accepted value for Cu at the diffusion-limited current density. For the magnetic layers, the same current efficiency was applied since according to previous profilometric measurements [16], the current efficiency is 96 % for bulk Ni-Co layers. Furthermore, recent XRD and TEM studies on ED Co-Cu/Cu multilayers [18,19] indicated that under such controlled deposition conditions, the actual layer thicknesses were, indeed, only slightly above the nominal values determined from the electrodeposition parameters.

The bilayer numbers were chosen so that the total thickness of all Ni-Co/Cu multilayers was about 300 nm and the nominal thicknesses of the individual layers were $d_{Cu}$ = 5 nm and $d_{NiCo}$ = 2 nm.

For purposes of comparison, two other sample series were also prepared, each with one single magnetic element in the magnetic layer (either Ni or Co). This was achieved by using only the required metal sulfate in the solution, while the overall metal ion concentration was the same as for the Ni-Co bath.

The overall multilayer composition was measured with electron probe microanalysis (EPMA) in a JEOL JSM-840 scanning electron microscope.

The root-mean-square surface roughness ($R_q$) of the deposited multilayers was determined by atomic force microscopy (AFM) with a Veeco Digital Instruments CP-II device. A white-light interferometer device of type ZYGO-LOT NewView 7100 was also used for measuring the surface roughness. The Si/Cr/Cu substrate showed height fluctuations not larger than 3 nm as established by both methods.

X-ray diffraction (XRD) was used to investigate the structure of the multilayers with the help of a Philips X'pert powder diffractometer in the $\Theta-2\Theta$ geometry with Cu-Kα radiation. Lorentzian curves were fitted on the background-corrected XRD diffractograms to determine the peak positions and the full width at half maximum (FWHM) of the peaks.

*2.2 Magnetoresistance measurements*

The room-temperature magnetoresistance (MR) was measured as a function of the external magnetic field (*H*) up to 8 kOe. The MR ratio was defined with the formula $MR(H) = (R_H - R_0)/R_0$ where $R_0$ is the resistance of the sample in zero external magnetic field and $R_H$ is the resistance in an external magnetic field *H*. The magnetoresistance was determined at room temperature in the field-in-plane/current-in-plane (FIP/CIP) geometry in both the longitudinal (LMR, magnetic field parallel to the current) and the transverse (TMR, field perpendicular to the current) configurations with a four-point-in-line probe. The measured *MR*(*H*) curves were decomposed, according to a procedure described previously [20] into ferromagnetic (FM) and superparamagnetic (SPM) contributions in order to determine the corresponding GMR terms $GMR_{FM}$ and $GMR_{SPM}$, respectively.

## 3. Results and discussion

*3.1 Influence of Cu deposition potential on multilayer composition*

Figure 1(a) displays a cyclic voltammogram for the electrolyte used for the Ni-Co/Cu multilayer deposition. Here and in all subsequent diagrams, the dashed vertical line indicates the value of $E_{Cu}^{EC}$ = -0.585 V, the electrochemically optimized Cu deposition potential for this particular bath which was determined according to the method suggested in Ref. 10.

Figure 1(b) shows the current transients during the P-pulse (Cu deposition). In this graph, the deviation of the current density from the steady-state value is presented in order to better reveal whether an anodic or cathodic transient occurs. The optimized $E_{Cu}^{EC}$ potential is characterized with the fastest but still positive current transient, which is mainly due to the interfacial capacitance of the metal/electrolyte interface.

As a next step, we present the composition analysis results for the investigated multilayers and discuss them in terms of the electrochemical reactions described in the Introduction. The overall composition of the multilayers is shown in Fig. 2 as a function of $E_{Cu}$ whereas Fig. 3 displays the relative Co and Ni concentrations in the magnetic layer. The actual layer thicknesses are also shown in Fig. 2 (right axis). It can be seen that the magnetic layer thickness varies from about 2.0 to 1.5 nm. By considering that the bilayer repeat length is about 7 nm, this small variation of the magentic layer thickness has a minor influence only on the GMR magnitude the value of wich is mainly determined by the rather large thickness of the Cu layer (about 5 nm).

Since Co and Ni are codeposited anomalously, the Co concentration in the deposit is always higher than the relative concentration of the $Co^{2+}$ ions in the electrolyte. Although the $Co_{50}Ni_{50}$ concentration in bulk alloy electrodeposits obtained by d.c. plating from the same solution [16] is reached at an ionic ratio $Co^{2+}/(Co^{2+} + Ni^{2+}) = 0.25$, the same deposit composition is reached at an ionic ratio of 0.05 in the magnetic layers of Ni-Co/Cu multilayers. The difference in the composition of the bulk and the nanometer-scale Ni-Co layers is explained by the long-lasting depletion effect that takes place in the electrolyte during the achievement of the steady-state deposition conditions. In this study, we present results on multilayers with a Co content in the magnetic layer slightly lower than 50 at.% (see Fig. 3).

If the Cu deposition potential $E_{Cu}$ is more positive than the electrochemically optimized value $E_{Cu}^{EC}$ = -0.585 V for this particular bath, Co starts to dissolve from the previously deposited magnetic layer. On this time scale, even for the most positive potentials, the dissolution of Ni can be considered as negligible in comparison with Co. This has two immediate consequences.

Firstly, the amount of charge driven through the system which is used for controlling the Cu layer thickness becomes the sum of two terms: the negative charge due to the deposition of Cu and the positive charge due to the dissolution of Co. Since the potentiostat measures the total charge only, the deposited Cu layer thus becomes thicker than expected from the preset charge. This conclusion is supported by the slightly increasing trend of the overall Cu content of the multilayers (Fig. 2) for values of $E_{Cu}$ potentials more positive than $E_{Cu}^{EC}$.

Secondly, for the potential range $E_\text{Cu} > E_{Cu}^{EC}$, the near-substrate region of the electrolyte becomes richer in $Co^{2+}$ than in the absence of the Co dissolution. Although it cannot be established whether the near-surface concentration of the $Co^{2+}$ ions indeed exceeds its "bulk" value in the electrolyte, it can be claimed that the $Co^{2+}$ ion concentration at the beginning of the next high-current pulse applied for the magnetic layer deposition is higher than during the application of $E_{Cu}^{EC}$ as the Cu deposition potential. This affects the deposition of the next magnetic layer. Therefore, there is an initial increase of the Co content in the magnetic layer at the expense of the Ni content as $E_\text{Cu}$ starts to become more positive than $E_{Cu}^{EC}$ (see Fig. 3).

As $E_\text{Cu}$ becomes more and more positive, the Co content in the magnetic layer should also start to decrease due to the stronger and stronger dissolution of Co atoms as it can, indeed, be seen in Fig. 3. With the increased dissolution rate of Co, the Cu deposition pulse is getting longer and longer. Therefore, the Co atoms dissolved from the previously deposited magnetic layer do not stay in the near-substrate region of the electrolyte. If they did so, they would easily be deposited into the next magnetic layer. The diffusion of $Co^{2+}$ ions into the solution acts toward equalizing the concentration everywhere in the bath, and hence the impact of the Co dissolution is damped. The mass transport effect at the increased time scale results in the observed decrease of the Co concentration of the multilayers with $E_\text{Cu}$ potentials in the range between -0.42 and -0.26 V (and probably for even more positive values).

On the contrary, for Cu deposition potentials more negative than $E_{Cu}^{EC}$, the magnetic atoms start to be codeposited with Cu. This codeposition process, especially at the beginning of the Cu pulse, can take place without any nucleation barrier since the actual surface at the end of the high-current pulse is rich in Ni and Co. This leads to a significant increase of the total concentration of Co and Ni in the multilayer (Fig. 2). The more negative the potential, the more magnetic material is codeposited with Cu.

The increase of the molar fraction of Ni and Co for $E_\text{Cu} < E_{Cu}^{EC}$ is accompanied by the decrease in the Co/(Co+Ni) ratio (see Fig. 3). Since the molar fraction of both Ni and Co is larger at -0.74 V that at -0.58 V, one can conclude that the deposition of both metal takes place in the P pulse, too. Since the deposition of the magnetic metals during the P pulse is not stopped, the deposition is getting closer to the d.c. plating. In other words, the $Co^{2+}$ concentration in the electrolyte near the cathode cannot relax as much as it could do without the deposition of the magnetic metals during the P pulse. This leads to a modified deposit composition also in the G pulse because the decrease in the Co deposition rate must be accompanied by the increase in the Ni deposition rate, having the total current density constant. Hence, the decrease in the Co/(Co+Ni) ratio is a consequence of the limited rate of $Co^{2+}$ transport in the electrolyte. This depletion effect comes from the same origin as the concentration gradient in bulk ED alloys where the Co concentration decreases with increasing thickness [21].

*3.2 Surface roughness behavior*

The cumulative roughening of electrodeposits gives rise to an ever increasing surface roughness [22]. This effect was investigated by several authors for different substrates and for various metallic deposits as well as multilayers and a short summary of these works has been given recently [23]. Surface scans of two samples with significantly different surface roughnesses can be seen in Figure 4 as measured by both AFM and white light interferometry. The root-mean-square surface roughnesses calculated from such measurements as a function of the preparation parameters are presented in Figure 5.

As it can be seen in Fig. 5(a), tuning the Cu deposition potential drastically changes the surface roughness ($R_q$) of the investigated Ni-Co/Cu multilayers. For Cu deposition potentials slightly more positive than $E_{Cu}^{EC}$, a significant smoothening of the surface sets in very abruptly and then the surface roughness remains practically constant. In this potential region, Co atoms are removed from the surface because of the Co dissolution. The interval of the Cu deposition potential where the surface remains fairly smooth is in good agreement with the occurrence of the dissolution process in the current transients (see Fig. 1(b)). On the contrary, for Cu deposition potentials more negative than $E_{Cu}^{EC}$, a dissolution of the magnetic layer does not occur but magnetic atoms start to codeposit with Cu. Due to the high deposition preference of Co when codeposited with Ni, the actual surface will be rich in Co.

Figure 5(a) also indicates that, in contrast to the Ni-Co/Cu multilayers, the surface roughness of Ni/Cu and Co/Cu multilayers does not vary significantly with Cu deposition potential: the data are scattered more or less randomly. This suggests that the observed roughening of Ni-Co/Cu multilayers for more negative potentials with respect to $E_{Cu}^{EC}$ is closely connected to the presence of both Ni and Co atoms at the deposit surface during the deposition period of the Cu layer.

To underpin the roughening effect due to the simultanous presence of $Ni^{2+}$ and $Co^{2+}$ ions, seven further multilayers were prepared at the most negative potential used (-0.74 V). Each of them had different relative Co concentration in the magnetic layer as presented in Fig. 5(b). The roughness values of the multilayers with either Ni or Co and the one with the $Ni_{50}Co_{50}$ alloy as magnetic layer coincide well with the values obtained on the previously discussed samples which were presented in Fig. 5(a). According to Fig. 5(b), the roughness values show a pronounced maximum with change in the composition of the magnetic layer. This means that this roughening does not come from the potential used but from the simultaneous codeposition of Ni and Co with Cu during the formation of the non-magnetic layer. The roughness reaches its maximum around a Co to Ni ratio of 1. Based on the composition of the sample prepared with the most negative potential in the P pulse (-0.74 V, see also Fig. 2), the overall Ni and Co content of the Cu layer can be assessed as 13 at.%.

A possible explanation for this increased roughening for anomalously codepositing metals can be the model described by Zech et al. [24] according to which the simultaneous deposition of Co and Ni may proceed not only in independent reaction routes as the case is for the deposition of Co or Ni alone in the absence of the other component. Instead, in the anomalous codeposition mechanism, the intermediates may contain the atoms of both metals, hence establishing a catalytic mechanism. The adsorbed reaction intermediates in the catalytic reaction may also contain several hydroxyl groups. Therefore, the adsorbed species, due to their large size, can effect the deposition of the majority component Cu as well. Due to the temporary blocking of some areas from growth by the intermediates, deposition will faster proceed at other places and this naturally leads to the observed surface roughening for the Ni-Co/Cu multilayers at large negative Cu deposition potentials. The locally and temporarily hindered simultanous deposition of the three metals excludes the possibility of the formation of larger islands of the same element which can lead to the observed roughening of the surface.

*3.3 X-ray diffraction studies*

An XRD study was carried out for all Ni-Co/Cu multilayers and the recorded XRD patterns are shown in Fig. 6 for the $2\Theta$ range from 40 to 54 deg. Further peaks of smaller intensity could also be detected at $2\Theta$ values around 74, 90 and 96 degrees. All detected peaks can be assigned to a face-centered cubic (fcc) structure whereas no peak was detected at $2\Theta = 41.68$ deg which is the typical peak appearing if a significant hexagonal close packed (hcp) phase is present in Co/Cu multilayers [18,25].

In the case of our samples, only the multilayers deposited with the most positive potentials showed satellite peaks (see Fig. 6) which prevailed up to the peaks at $2\Theta = 90$ degrees. From the positions of the main and the satellite peaks for the (111) reflection, the bilayer thicknesses ($\Lambda$) were calculated. Since only the intensity of the left-side satellites were high enough for a meaningful fitting, this limited the accuracy of the bilayer period determination. For $E_{Cu}$ potentials -0.42 V, -0.34 V and -0.26 V, the experimental values for $\Lambda$ were 7.8 nm, 7.9 nm and 8.2 nm, respectively. These values are slightly larger than the nominal value ($\Lambda = 7.0$ nm) and this is in agreement with the results of previous studies on ED Co/Cu and Ni/Cu multilayers [9,18,25,27,28].

A comparison of Figs. 5 and 6 reveals that satellites could only be observed for multilayers with sufficiently smooth surfaces. This is in agreement with expectation since for rough surfaces the degree of structural coherence, a pre-requisite for the appearance of satellites [26], definitely decreases.

At the same time, the FWHM data were also in agreement with the conclusions from the observation of the satellite peaks. Namely, the XRD line broadening of both the main peaks and the satellites as measured by FWHM correlates well with the surface roughness. The linewidth was found to be smaller for the more positive potentials and this indicates that a low surface roughness of the multilayers is accompanied by a higher degree of structural perfectness. This is a result of the dissolution-induced smoothening of the magnetic layer at more positive potentials than the electrochemically optimized $E_{Cu}^{EC}$ value. The narrowing of the satellite peaks (and the simultaneous increase of their intensity) along the decreasing surface roughness is particularly informative. Namely, since a fluctuation of the thicknesses of the individual layers results in a broadening of the satellite peaks [29], the observed continuous decrease of the FWHM of the satellites confirmed that the thickness fluctuation of the layered structure was diminished by the dissolution-induced smoothening of the magnetic layer.

An attempt was also made to deduce an effective lattice constant from the positions of the XRD peaks even if we know that there is a tetragonal distortion of the cubic cells due to the in-plane lattice mismatch of the constituent layer elements. The peak position of reflection (hkl) determines the average lattice plane distance $d_{hkl}$. By using the relation $1/d_{hkl}^2 = (h^2 + k^2 + l^2)/a^2$ [30], an effective lattice constant $a_{hkl}$ can be deduced for each reflection observed. The resulting experimental $a_{hkl}$ data are shown in Fig. 7. The $a_{hkl}$ values deduced for the three main reflections coincide well for a given sample. Although there is a large scatter of the $a_{hkl}$ values which prevents to reveal a particular dependence on $E_{Cu}$, at least these values are close to the expected ones. By using Vegard's law, Fig. 7 shows the lattice parameters for bulk fcc-Co and fcc-Ni as well as for a $Ni_{50}Co_{50}$ alloy. This value is then interpolated again with Cu via Vegard's law to arrive at the expected lattice constant for an ideal $Ni_{50}Co_{50}$/Cu multilayer (all these latter data are represented by horizontal straight lines in Fig. 7. The two solid lines in Fig. 7 represent the effective lattice parameters by taking into account the results of chemical composition analysis and Cu layer thickness variation as deduced from the composition data for the various Cu deposition potentials. The experimental $a_{hkl}$ data for the $Ni_{50}Co_{50}$/Cu multilayers are close to the estimated lattice parameters and even the evolution of the two sets of data with $E_{Cu}$ can be considered as satisfactory. Thus, Fig. 7 evidences that the nominal layer thicknesses, the measured multilayer compositions and the XRD results constitute a consistent set of data for the multilayers investigated.

*3.4 Giant magnetoresistance*

All the Ni-Co/Cu multilayers investigated exhibited GMR at room temperature in that both the LMR and TMR components were negative for the whole range of magnetic field applied. The $MR(H)$ curves nearly reached complete saturation in magnetic fields around 2 kOe after which the resistivity change was already much smaller as shown for a typical multilayer in Fig. 8. The saturating component is due to a ferromagnetic contribution to the GMR whereas the slowly saturating component is due to the presence of a small amount of superparamagnetic regions in the magnetic layers [20].

After separating the two contributions by a standard procedure [20] (as indicated for the TMR component in Fig. 8), Fig. 9 shows the ferromagnetic contribution $GMR_{FM}$ as a function of the Cu deposition potential and the ratio of the superparamagnetic GMR contribution GMR ($GMR_{SPM}$) to the saturation value of the GMR ($GMR_S$) as a funtion of the Cu deposition potential. Since the difference between the LMR and the TMR values was small, an isotropic average of the GMR values was only displayed.

The $GMR_{FM}$ contribution has a maximum at about -0.4 V which is the result of several different but not independent effects. The root-mean-square surface roughness has evidently a significant effect on the $GMR_{FM}$ contribution (see Fig. 10, left ordinate): the smoother the multilayer surface, the larger the $GMR_{FM}$ contribution. This should be connected with the improvement of the structural quality of the multilayers for $E_{Cu}$ values more positive than $E_{Cu}^{EC}$ as was revealed by the XRD data above.

For very negative Cu deposition potentials, magnetic atoms are codeposited in the Cu layer (see Fig. 2). The presence of magnetic Co and/or Ni atoms in the Cu layer can effectively decrease the spin diffusion length of the electrons in the spacer layer even at moderate concentrations, which strongly diminishes the GMR effect. At large concentration of the magnetic atoms, especially when they are enriched locally, they can form a direct magnetic connection between adjacent magnetic layers and the resulting ferromagnetic coupling also reduces the GMR.

As noticed above, some fraction of the observed GMR arises due to the presence of SPM regions in the magnetic layers. In Fig. 10 (right ordinate), the ratio of the superparamagnetic contribution ($GMR_{SPM}$) to the saturation value of the GMR ($GMR_S$) is plotted as a function of the surface roughness. A clear correlation can be observed in that the SPM contribution to the GMR increases with increasing surface roughness. Ishiji and Hashizume [31] reported similar results in that for sputtered Co/Cu multilayers the $GMR_{SPM}$ contribution was found to increase with substrate roughness. These authors have also put forward a model how a rough surface can lead to the appearance of isolated magnetic regions during multilayer growth. For our ED multilayer samples, the correlation between roughness and the SPM contribution comes about via the Cu deposition potential (see Fig. 9). When changing $E_{Cu}$ towards more positive values, a surface smoothening occurs as a result of the dissolution of all leachable magnetic material including also regions with SPM behavior. At the most positive Cu deposition potentials, there is still a small residual $GMR_{SPM}$ component (about one tenth of the total GMR, see Fig. 9). This derives from the remaining superparamagnetic regions of the magnetic layers which are not accessible for the dissolution process and thus when tuning the Cu deposition potential to more positive values, this ratio does not change significantly in agreement with the saturation of the surface roughness values for the Ni-Co/Cu multilayers (see Fig. 5(a)).

## 4. Conclusions

In the present work, the effect of the Cu deposition potential on the microstructure and GMR was investigated for ED Ni-Co/Cu multilayers. The Cu deposition potential was varied in a wide range around the electrochemically optimized potential $E_{Cu}^{EC}$ which corresponds to deposition conditions ensuring that no unwanted electrochemical reactions take place during the deposition pulse of the more noble non-magnetic element.

The surface roughness study with both AFM and interferometry revealed that the roughness changes radically for Ni-Co/Cu multilayers as $E_{Cu}$ is tuned away from the optimal $E_{Cu}^{EC}$ value. For more negative potentials, the roughness increases and for more positive potentials, the roughness decreases first and then remains constant. As a comparison, the surface roughness evolution with $E_{Cu}$ was investigated also for ED Co/Cu and Ni/Cu multilayers and $R_q$ was found to remain nearly constant for the same Cu deposition potential range. Surface roughness data on an additional Ni-Co/Cu multilayer series for which the magnetic layer composition varied from pure Ni to pure Co indicated a pronounced roughness maximum around a Co to Ni ratio of 1. The different roughening behavior for the binary and ternary systems could be qualitatively explained with the help of the catalytic codeposition mechanism in the case of Ni-Co alloys which was suggested in the literature previously [24].

According to an XRD study, all the Ni-Co/Cu multilayers exhibited an fcc phase. In conformity with the better structural quality indicated by the smoother surface for sufficiently positive Cu deposition potentials, the same samples showed satellite reflections as well. From the satellite positions, the bilayer repeat period was deduced to be only slightly larger (by about 10 to 20 %) than the nominal values, in agreement with previous observations on ED multilayers. From the positions of the XRD peaks, an effective lattice constant could be derived which were in good agreement with the values estimated from the pure metal values by using Vegard's law.

The magnetoresistance of the multilayers showed a maximum at $E_{Cu}$ = -0.42 V which is more positive than the electrochemically optimized value $E_{Cu}^{EC}$ = -0.585 V. This is in the Cu deposition potential range where both surface roughness and XRD data indicated a better structural quality of the multilayer. Thus, the present study points out that surface roughness also plays an important role in determining the GMR in ED multilayers.

**Acknowledgements**

This work was supported by the Hungarian Scientific Research Fund through grant OTKA K 75008.

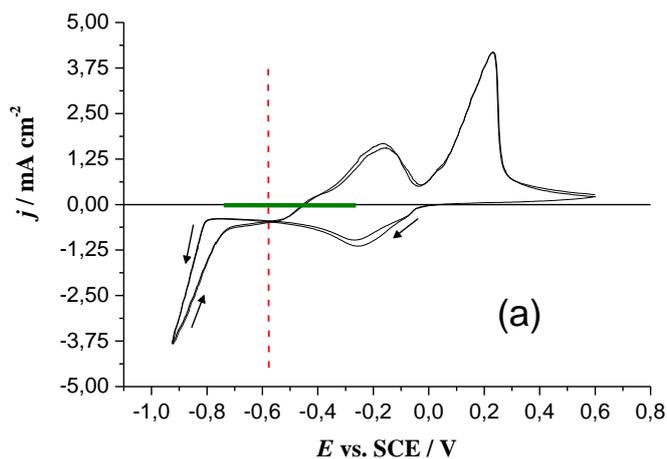

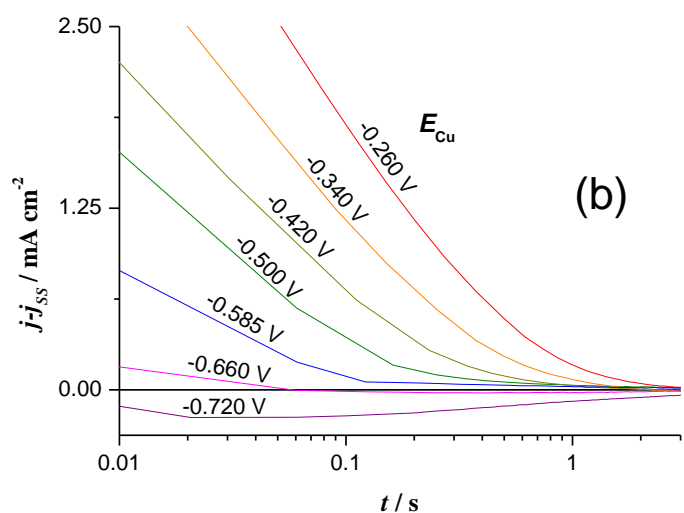

*Fig. 1* (color online only) (a) Cyclic voltammogram of the electrolyte used in the present work for the electrodeposition of Ni-Co/Cu multilayers. The vertical dashed line (here and in all subsequent figures) indicates the value of the electrochemically optimized $E_{Cu}^{EC}$ potential for the deposition of the Cu layer. The thick horizontal line shows the potential range investigated in this paper. (b) Current transients for the Cu deposition pulse as a function of the time after the start of the same pulse. The curves were corrected by deducting the steady-state current density in the same pulse ($j_{SS}$) in order to highlight the sign of the deviation from the steady-state current density in the early phase of the P pulse.

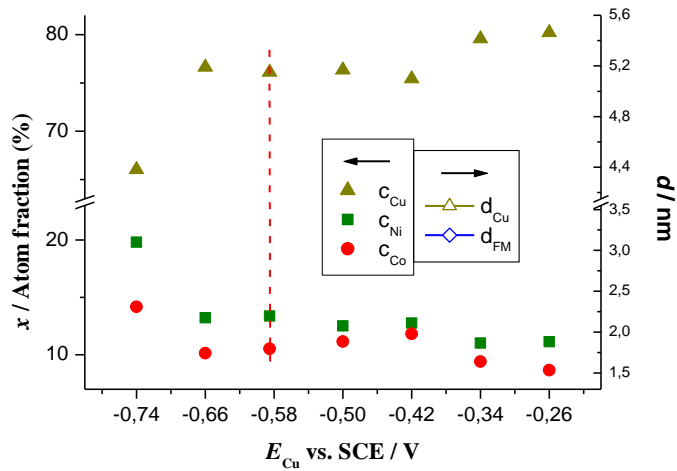

*Fig. 2* (color online only) Dependence of the overall composition of the multilayers on the Cu-deposition potential. In this figure, $x(Ni)+x(Co)+x(Cu) = 100$ %. The full symbols represent the concentration values of the multilayers (left axis) and the open symbols represent the calculated layer thicknesses (right axis) due to the exchange reaction on the basis of overall multilayer composition analysis data.

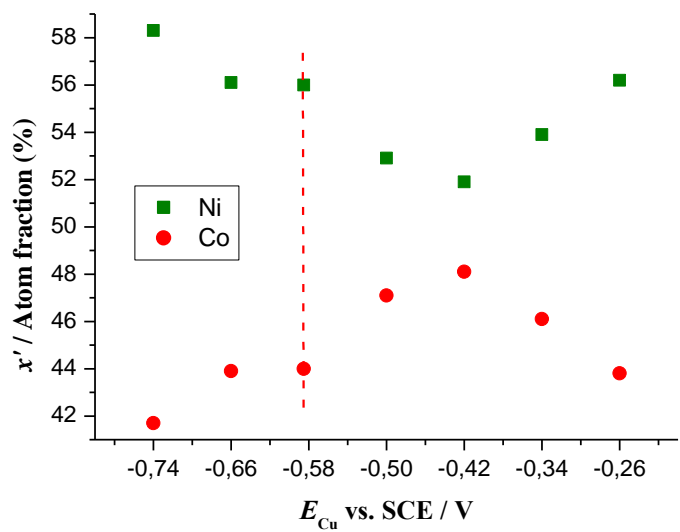

*Fig. 3* (color online only) Dependence of the composition of the magnetic layer on Cu-deposition potential. The small amount (ca. 1 at.%) of Cu codeposited in the magnetic layer is neglected here. Note that in this figure $x'(Co) + x'(Ni) = 100$ %.

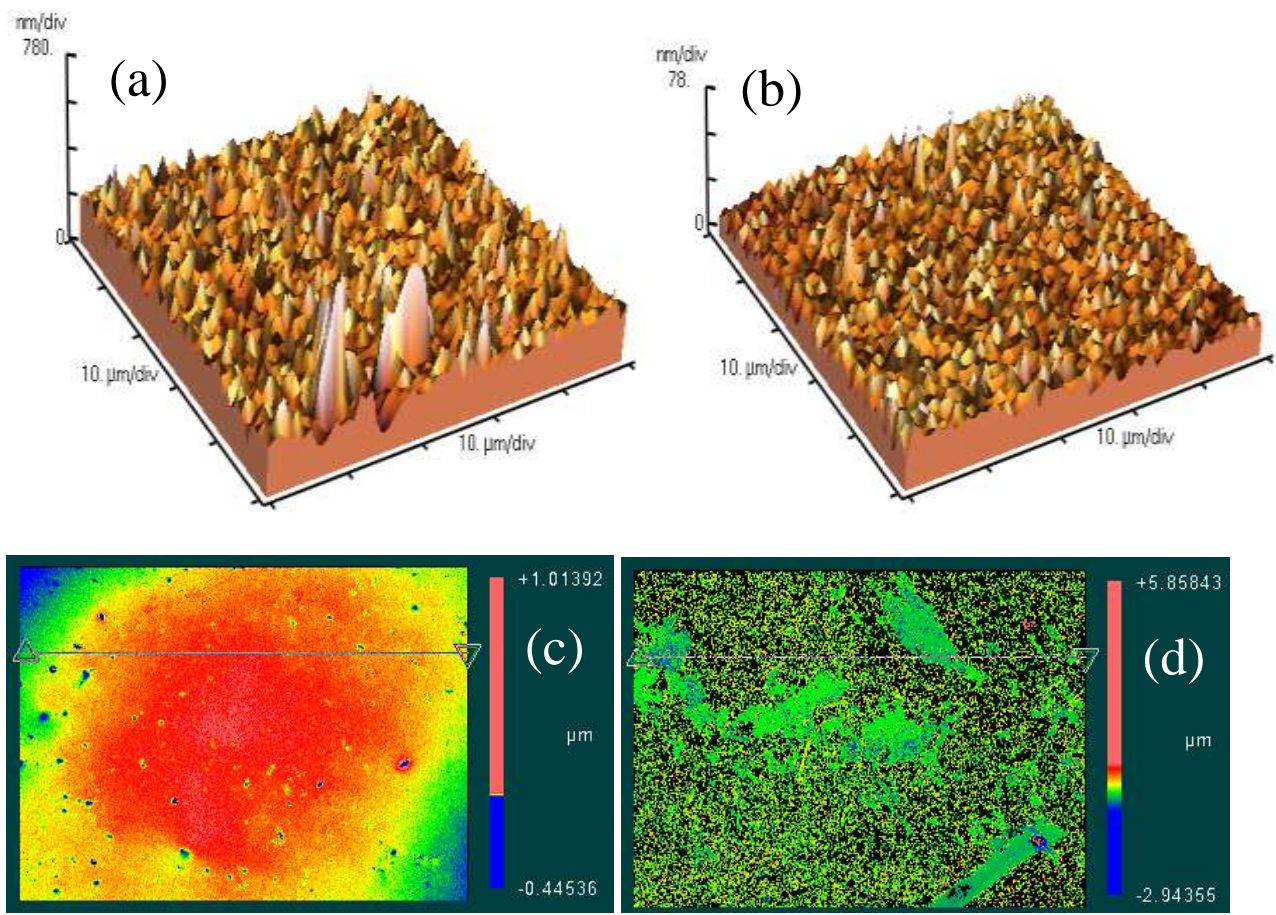

*Fig. 4* (color online only) Images of the surface of electrodeposited Ni-Co/Cu multilayers prepared at two Cu deposition potentials ($E_{Cu}$). (a) AFM image, $E_{Cu}$ = -0.74 V; (b) AFM image, $E_{Cu}$ = -0.26 V (note the one order of magnitude difference on the height scale); (c) interferometry image, $E_{Cu}$ = -0.74 V; (d) interferometry image, $E_{Cu}$ = -0.26 V. The $R_q$ parameter was evaluated on a 50 μm x 50 μm area for the AFM image and on a 0.94 mm x 0.70 mm are for the interferometry image.

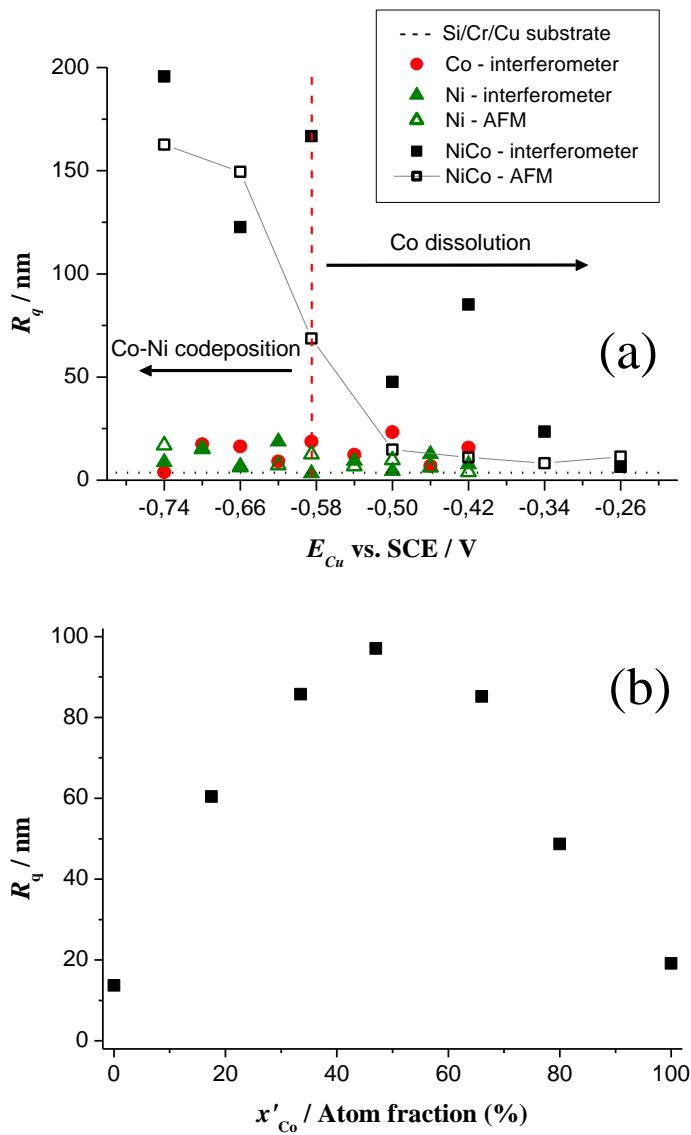

*Fig. 5* (color online only) (a) Dependence of the root-mean-square roughness $R_q$ on the Cu-deposition potential. At the electrochemically optimized Cu deposition potential $E_{Cu}^{EC}$ indicated by the dashed vertical line, neither Co and Ni dissolution, nor Co and Ni codeposition occurs. The full symbols represent the data measured with the white-light interferometer, the open symbols stand for the AFM data and the dotted line is the roughness of the substrate measured with both methods. (b) Dependence of the root-mean-square roughness $R_q$ on the Ni/Co ratio of the magnetic layer in the case of $E_{Cu}$ = -0.74 V.

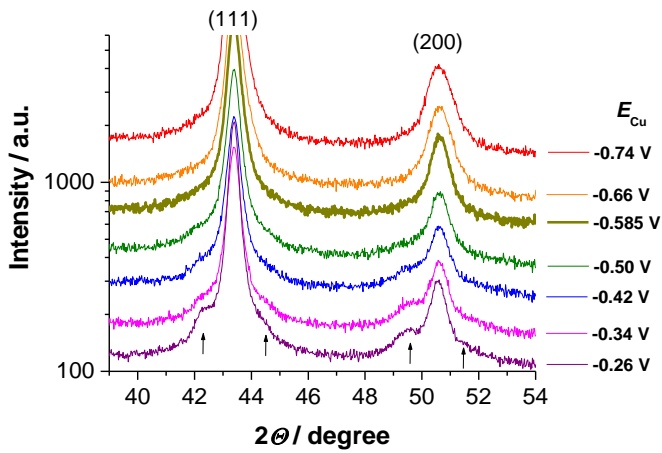

*Fig. 6* (color online only) X-ray diffractograms of the investigated Ni-Co/Cu multilayers. The arrows indicate the approximate position of the satellites for the sample deposited at $E_{Cu}$ = -0.26 V.

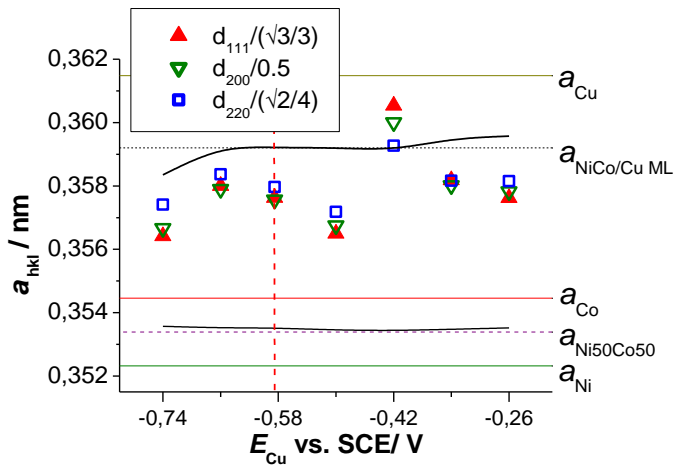

*Fig. 7* (color online only) Calculated effective lattice parameters for the $E_{Cu}$ values investigated. The data points are derived from the XRD peak positions. The curves and dashed horizontal lines are calculations from Vegard's law. The three horizontal solid lines refer to the three constituent metals (Ni, Co, Cu). The lower dashed line is for a $Ni_{50}Co_{50}$ alloy. The upper dashed line is for the whole multilayer calculated by assuming a 2 nm thick $Ni_{50}Co_{50}$ alloy as magnetic layer and 5 nm Cu as non-magnetic layer. The lower thick solid curve is for the magnetic layer calculated from the measured Co and Ni concentrations. The upper thick solid curve is for the whole multilayer sample when taking into account the actual concentrations of the metals in the multilayer and the Cu layer thickness calculated from the measured concentrations by assuming 7 nm bilayer thickness.

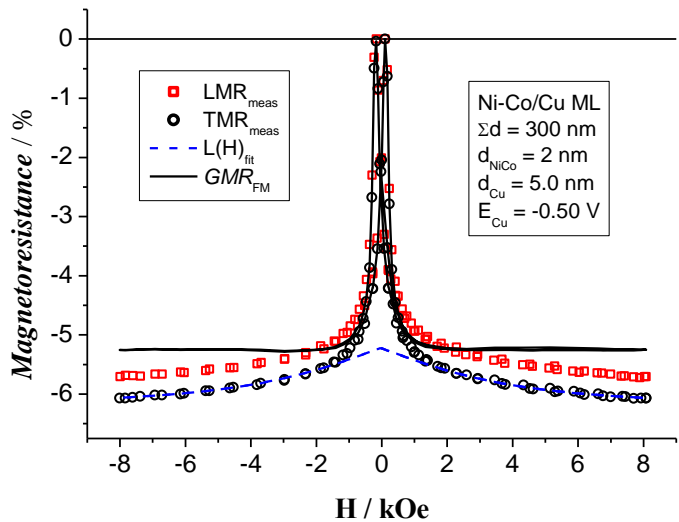

*Fig. 8* (color online only) A typical measured *MR*(*H*) curve. The open symbols represent the measured MR values (□: LMR, ○: TMR), the dashed line is the Langevin function fitted to the measured TMR data and the solid line is the FM contribution to the GMR which can be obtained by subtracting the fitted Langevin function from the measured values. The dashed line actually represents the field evolution of the SPM contribution to the GMR.

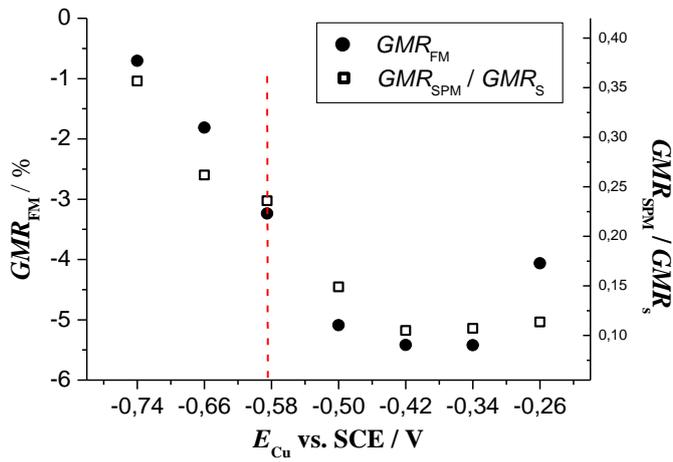

*Fig. 9* (color online only) Dependence of the ferromagnetic contribution *GMR*<sub>FM</sub> (solid symbols) and the ratio of the superparamagnetic contribution of the GMR to the saturation value of the GMR (open symbols) on the Cu deposition potential.

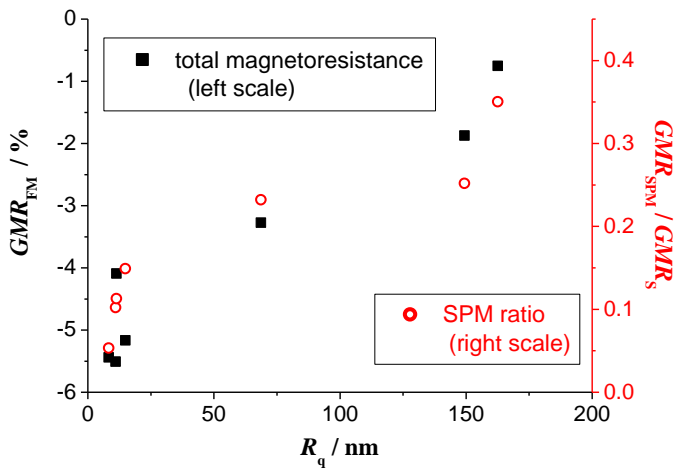

*Fig. 10* (color online only) The observed ferromagnetic contribution to the GMR and the fractional superparamagnetic contribution to the saturation value of the GMR as a function of the root-mean-square surface roughness.